\documentclass[%
 preprint,
 amsmath,amssymb,
 aps,
]{revtex4-2}
\usepackage{graphicx}

\usepackage[mathlines]{lineno}
\modulolinenumbers[5]
\linenumbers\relax 

\usepackage{graphicx}
\usepackage{amssymb}
\usepackage{amsthm}
\usepackage{amsmath}
\usepackage{amsfonts}
\usepackage{bm}
\usepackage{xcolor}
\usepackage{comment}
\usepackage[export]{adjustbox}

\newcommand{\beq}{\begin{equation}}
\newcommand{\eeq}{\end{equation}}
\newcommand{\ba}{\begin{eqnarray}}
\newcommand{\ea}{\end{eqnarray}}

\newcommand{\R}{I\kern-0.3emR}
\newcommand{\E}{\mathbb{E}}

\begin{document}

\preprint{}
\title[Societies of Neural Networks]{Frustration, glassy behavior and dynamical annealing    in societies of Neural Networks.}

\author{Felippe Alves}
 \email{flipgm@gmail.com}
\author{Nestor Caticha}%
 \email{ncaticha@usp.br}
\affiliation{Instituto de Fisica, \\Universidade de Sao Paulo,\\
CEP 05315-970, S\~ao Paulo, SP, Brazil}%

\date{\today}

\begin{abstract}
We study maximum entropy mechanisms of information exchange between agents modeled by neural networks and the macroscopic states of a society of such agents in a few situations. 
Mathematical quantification of surprise, distrust of other agents and confidence about its opinion emerge as essential ingredients in the entropy based learning dynamics. Learning is shown to be driven by surprises, i.e. the receptor agent is confronted with the concurring opinion of a distrusted agent or with a trusted agent's disagreeing opinion. Attribution of blame for the surprise derives from measures of distrust of the
receiver towards the emitter agent and the receiver's confidence about its own opinion. The dynamics proceeds by changes of mainly one or the other: the receptor opinion about the issue or the distrust about the emitter.
A society with $N$ agents exchanging  binary opinions about a set of issues show rich behavior which depend on the complexity of the agenda. For small sets the society reaches a steady state polarized into antagonistic factions, where balanced norms such as ``the friend of an enemy is an enemy" are strictly satisfied.
For larger sets of issues, societies can persist for a long time in spin-glass like states. There are two types of frustration: ideological and affective, with dynamical annealing properties depending on the complexity of the set of  questions under discussion, leading to the lack of sharply defined parties for long transients. 
\end{abstract}


\maketitle

\section{Introduction}
The analysis of opinion polarization in societies is central in several areas, see  \cite{reviewAgentsIEEE} for a review of the literature.  
Agent based modeling has has been used to build interacting systems that show collective properties of antagonistic bands. Some models \cite{Axelrod1997}  include the multivariate representation of the cultural state of an agent and the information  exchanges with neighbors  on a given geographical setting. Alternatively,
 agents can be Ising like variables, i.e. represented by the two states of ``for" or ``against" opinions on a given single issue
\cite{Galam08}\cite{Castellano09} which, with different combinations of ferromagnetic and  anti-ferromagnetic interactions  yields interesting collective behaviors.
Even with these simplifications of the agents,  several empirical successes can be obtained in  complex situations. Agents may have continuous or vector opinions \cite{Deffuant2000} or have simple Bayesian learning \cite{Martins_2009}. 

We follow a different modeling path, using agents that are simple neural networks (NN) with psychologically inspired traits and rich learning dynamics, as recommend by psychologists \cite{jagerEROS2017}.
Our agents exchange information in a way that saturates bounds of efficiency in different student-teacher scenarios. Information theory and Statistical Mechanics guide the development of a framework, the entropic dynamics for NNs algorithm (EDNNA) \cite{CatichaEdnna2020}, for general architectures. The resulting learning algorithms reveal  aggregate  variables  identifiable  with correlates in humans, such as  cognitive dissonance, surprises, performance, the distrust of an individual and confidence on an opinion, that characterize the interactions of a  pair of agents.  Macroscopically, frustration, social balance, its emergence or breakdown and  different types of affinities, e.g. affective or ideological \cite{Jost2009a} can drive polarization and lead to different types of spin-glass like states.

Adaptive changes through learning is the basic assumption of this work. Therefore, we choose the NN of simplest architecture, the single layer perceptron, as a model for each agent.
See \cite{Metzler00} for early work on interacting perceptrons. Linearly separable models have been shown to be useful in describing human performance in several cases \cite{RescorlaWagner}. The resulting societies will be shown to be already sufficiently complex that at least, this step is justified as necessary in the path towards richer agent models.

There are several learning algorithms for a perceptron, but we can imagine a process of evolution, where the selection pressure comes from the needs of better generalization ability. This has been studied in several papers analytically \cite{KiCa1992}, \cite{CoCa1995}, \cite{Opper96},\cite{CaKi97Time}, \cite{Opper99}, \cite{Solla98}, \cite{CatichaEdnna2020} and using evolutionary programming \cite{Neirotti2003},\cite{CaNe2006ToBayes}. Different simpler versions have already been applied to societies \cite{ViMaCa2009},\cite{CaVi2011},
\cite{CaCeVi2015} and \cite{alves2016sympatric}. Data from
decisions in 3 judges panels of US appellate courts has been studied \cite{CatAlvEsann2019} with these models. Other societies of NN have been studied by \cite{Neirotti16} and 
\cite{Neirotti17}. 

While selection of psychological traits, or analogously optimization of learning algorithms can be performed under certain conditions, such as in small groups, people and agents are called upon to engage in social interactions in groups far larger than those of the original environment. Optimality depends on the context and as the agents here studied, we have not evolved to deal within the modern contexts of large groups \cite{Dunbar1992}. 

The performance of a NN learning a classification  task from examples, will suffer from noise affecting the learning set. Introducing the possibility of noise among the assumptions used to optimize a learning rule enhances the machine's performance.  A measure of  how much a receiver should distrust the  emitter's signal is equivalent to a noise level estimate and entropic inference determines the rule for dynamical update of distrust. Earlier attempts to introduce adaptive noise estimation for a single agent with this architecture include \cite{biehl1995noisy}, \cite{copelli1997noise}. Communication can also be disrupted by parsing errors which would introduce noise to the mathematical representation of an assertion under discussion.

The society evolves under a discrete time dynamics. At a given time, two agents are selected to interact and randomly assigned the emitter and receiver roles. A question is chosen from the prominent questions of policy which occupy the public attention. The emitter sends its opinion. The receiver changes its state using the learning algorithm. There are 
several conceivable ways to choose  the pair of agents. Borrowing language from biology, \cite{alves2016sympatric} use the term {\it allopatric} group formation when the probability of choosing an emitter depends on the receiver's distrust. An effective communication barrier is created by repeated disagreement and this pair of agents cease to interact, grouping into  factions holding opposing opinions.  Alternatively,  {\it sympatric} describes the process where agents keep interacting despite holding  opposing views on the set of issues. We show that this dynamics of distrust  permits  the possibility of anti-learning, i.e learning  the reverse of the emitter's opinion.

In section \ref{dynamics}, we present the technical aspects of learning by entropic dynamics in general settings. This theory is applied in section \ref{ideology} to agents that have  two sectors, for ideological/opinion and affective/distrust parameters, respectively, the weights of the NN and the noise level estimate of the channel. In section \ref{algorithmdescription} we show that  learning occurs with high intensity when there is a surprise: the receiver agrees with a distrusted emitter or disagrees with a trusted one. A change occurs in the receptor to decrease this dissonant interaction. The receiver's changes resemble the learning algorithms for a tree committee machine \cite{CoCa1995} and a parity machine \cite{SimCa96}. These Bayesian algorithms are poorly approximated by the least action introduced by \cite{Kabashima} for the parity machine and \cite{MitDurb} by hand. There is an interaction between the distrust and the weight sectors. The largest change occurs in the sector in which the agent is the least confident. Learning proceeds by predominantly changing one sector for the surprise  such that the dissonance decreases.

In this large $N$  society model, section \ref{Nagents}, there is no absolute truth since there is no objective external fact checking mechanism. As Lincoln  \cite{Lincoln} already knew
``The process is this: Three, four or half a dozen questions are prominent at a given time, the party selects its candidate, and he takes his position on each of these questions.''
If only a few assertions are under discussion i.e. the prominent issues, the system typically evolves into two separate parties. Membership can be attributed in two ways, by distrust or opinions. Distrust induced polarization  is analogous to affective polarization  discussed in political science and opinion polarization to ideological polarization on the basis of beliefs about policies. Scholarship \cite{KlarAffective}\cite{IyengarAffective} about their theoretical characterization and  experimental  measurement is rapidly accumulating. 

Frustration, \cite{AndersonFrustration} \cite{ToulouseFrust}, also comes in two flavors. One similar to the structural balance theory  \cite{HarariBalance} in social psychology is affective frustration. 
 Agents $I,J$ and $M$  are in a balanced relation if the distrusts of $I$ towards $J$, from $J$ towards $M$ and from $M$ towards $I$ contains an even number of antagonistic relations. The unbalanced case of an odd number of antagonistic  relations is frustrated and describes triplets of agents where for example ``the friend of a friend is not a friend."  The second type,  ideological frustration, describes unbalanced ideological alignment. As the number of issues under discussion becomes larger the society resembles more a spin-glass, where the number of frustrated triples is macroscopically large for persistently longer times. An interesting question in current affairs is to establish whether ideological polarization leads or follows affective polarization. We find that for agents with the full EDNNA discussing a simple agenda, affective polarization is established faster and drives ideological polarization. As the set of issues  gets larger this is reversed and ideological polarization drives affective polarization. For a simpler learning algorithm, this reversal doesn't occur.

\section{Entropic Learning Dynamics\label{dynamics}}
In online learning, an example input-output pair $(\bm x_t, \sigma_e) $ drives the dynamics one  time step. 
In Bayesian learning \cite{Opper96}, the likelihood $L$ is a probability that incorporates the information about the architecture of the networks and pushes the prior to a posterior distribution of the NN parameters. 
We choose to describe, at time $t$, the prior by  a member of a MaxEnt parametric family    of densities such that the knowledge about the parameters $\bm u$ of the network, conditional on the data prior to $t+1$,  is represented    by $Q_{t}$
\beq 
Q_{t}(\bm u |D_{t}) = \frac{1}{\zeta}e^{-\sum_a \lambda_{a, t} f_a(\bm u)},
\eeq 
where the set of generator functions $\{f_a(\bm u)\}$ defines the manifold; $D_{t}=\{\lambda_a\}$ is the set of Lagrange multipliers that enforce a set of constraints   ${\cal F}_a^{t}= \E_{Q_{t}}(f_a)$ and $\zeta$  ensures normalization. Bayes update is given by
\beq
P_{t+1}: =P(\bm u |\bm x_t, \sigma_e, D_{t}) = \frac{Q_{t}(\bm u|D_{t}) L_t(\sigma_e| \bm x_t, \bm u, D_{t})}{Z(\sigma_e|\bm x_t, D_{t})},
\eeq 
where the the evidence for the model is the probability $Z_t: =Z(\sigma_e|\bm x_t, D_{t}) =\int  L_{t}   Q_{t}d\bm u$. The likelihood doesn't depend on $D_t$, only on $(\sigma_e,\bm x_t)$ and the state of the network $\bm u$.
In general the Bayes posterior will not be in the manifold, i.e not conjugated. Nevertheless, the expected values  of the Maximum Entropy generator functions of the family, under the Bayes posterior, 
point uniquely to a new distribution ${Q_{t+1}}$ in the manifold.
This MaxEnt posterior is the new representation of knowledge for the network and the prior for the next step, obtained by maximizing
\ba
S[Q_{t+1}||Q_{t}]&=& -\int  Q_{t+1} \log \frac{Q_{t+1}}{Q_{t}}d\bm u \nonumber \\
&-&\Delta \lambda_a \left( \E_{Q_{t+1}}(f_a) -\E_{P_{t+1}}(f_a)\right),
\ea 
subject to the  constraints that its expected values $ \E_{Q_{t+1}}(f_a)$ are equal to the Bayes posterior expected values $\E_{P_{t+1}}(f_a)$. The Lagrange multipliers are denoted by $\Delta \lambda_a $ since they are the increments of the $\{\lambda_a\}$ of the prior $Q_{t}$. 
It follows that 
\ba
{\cal F}_a^{t+1}-{\cal F}_a^{t}
&=& -\frac{\partial {\log Z_t}}{\partial \lambda_{a, t}}, 
\label{entropicdyn}
\ea
which makes no mention of the Bayes posterior. These equations hold for any well behaved family.
Since this learning dynamics is deduced from entropy maximization it is called Entropic dynamics. 
Learning occurs along the gradient of the log evidence. An element of the MaxEnt manifold can be represented  equivalently by different sets of coordinates, the constraints ${\cal F}_a$ or the Lagrange multipliers $\lambda_a$ or any Routhian mixture  of the Legendre transforms. Convenience dictates which representation should be used.  
 We are interested here in the case  of linear and quadratic generators: $f_0=1$, $f_i = u_i$ and $f_{ij}=u_iu_j$, 
the Gaussian family, which we write in the conventional way
$Q_{t+1}\propto \exp [-\frac{1}{2} (\bm u- \hat{\bm u_t})\cdot \bm \Sigma_t^{-1}\cdot (\bm u- \hat{\bm u_t})]$.
The entropic dynamics update equations \ref{entropicdyn}, driven by the arrival of the $t^{th}$  example describe the changes in the parameters of $Q$, its mean $\hat {\bm u}_{t}$ and covariance matrix $\bm \Sigma_{t}$:
\ba
\hat {\bm u}_{t+1} &=&  \hat {\bm u}_{t} +\bm \Sigma_{t} \nabla_{\hat{\bm u}_{t}}   {\log Z}_{t}, \label{dinawvet} \\
\bm \Sigma_{t+1}
 &=&\bm \Sigma_{t} + \bm\Sigma_{t}(\nabla_{\hat {\bm u}_{t}}\nabla_{\hat {\bm u}_{t}}^{T}{\log Z_{t}}){\bm \Sigma}_{t}
\label{dinaCvet}.
\ea  
These were first obtained by Opper \cite{Opper96} using a different  argument, based on minimizing the expected value of a log loss function, which  turns out to be equivalent to the (less intuitive) maximization of $S[P||Q_{t+1}]$, the entropy of the Bayes posterior relative to a ``prior" $Q_{t+1}$, which is  the entropic posterior. 


\subsection{Ideological and affective sectors \label{ideology}}
Equations \ref{dinawvet} and \ref{dinaCvet} are general, they implement learning for any architecture and therefore to move on, we have to make some structural hypothesis.
We split the degrees of freedom $\bm u$ into two  subspaces, the ideological sector $\bm w \in \R^K$ and the affective sector $z \in \R$. Given the internal state of an agent, $\bm w$, the forced decision of being for or against an issue $\bm x$ is given by $\sigma  = \text{sign} (\bm w\cdot \bm x)$.

The affective sector models the distrust that the receiver agent has for the emitter agent. It could be higher dimensional, but here we restrict to one dimension, because we model it as a noisy channel, which has been studied  \cite{biehl1995noisy} \cite{KiCa1993} for perceptrons learning from examples in the student-teacher scenario. Information about the distrust $z$ of the receiver towards the emitter is encoded in a Gaussian with mean $\mu_{e|r}$ and variance $V_{e|r}$.

Although the receiver distrusts the emitter $\sigma_e$ is useful information about the putative ``true label" $\sigma_T$. Another source of noise may act on the issues themselves,  the issue $\bm x_t$ considered by the receiver is not $\bm y_t$, considered by the emitter, a ``parsing error". We take the components of $\bm x_t$ and $\bm y_t$ to be Gaussian correlated variables, $\E x_{t,i}=\E y_{t,i}= 0$, $\E x_{t,i} y_{t,i} = v_p$.  To obtain the likelihood we  marginalize the joint distribution of labels and the issue $\bm y_t$:

\ba 
L(\sigma_e|\bm x_t, \bm u)&=& \sum_{\sigma_T}\int    P(\sigma_e| \sigma_T, \bm y_t,\bm x_t, \bm u)P(\sigma_T|\bm y_t, \bm x_t, \bm u) \nonumber \\
& \,& P(\bm y_t|\bm x_t, \bm u) d\bm y_t.
\ea
Furthermore, $\sigma_T=$ sign$(\bm y_t \cdot \bm w_e)$ for an unknown $\bm w_e$ but it is reasonable to suppose that $\sigma_e$ depends on $\bm  y_t$ and $\bm w_e$ only through $\sigma_T$, so 
\ba
P(\sigma_e| \sigma_T, \bm y_t ,\bm x_t, \bm u,\bm w_e) &=&
P(\sigma_e| \sigma_T, z) \\
&=& (1-\varepsilon(z)) \delta_{\sigma_e,\sigma_T}+\varepsilon(z) \delta_{\sigma_e,-\sigma_T},
\nonumber \ea
where $\varepsilon(z):\R \rightarrow{[0,1]}$ is a function of $z$. 
This random variable, limited to the unit interval, is the probability of the emitter conveying $\sigma_e \neq \sigma_T$ to the receiver.
While tempting, the choice of a Beta distribution for $\varepsilon$ results in inconvenient technical difficulties.
With 
$\varepsilon(z) = \Phi(z)$, the  cumulative distribution of a standard  Gaussian and $z$ a Gaussian with mean  $\mu_{e|r}$ and standard deviation $V_{e|r}$, which are a representation of the distrust of the receiver and its uncertainty about it, with regard to the emitter agent, we maintain the Gaussian simplicity of equations \ref{dinawvet} and \ref{dinaCvet}.

Choosing a Gaussian $G(\bm y_t|\bm x_t, v_p(\bm w))$ for $P(\bm y_t,|\bm x_t, \bm u)$ and
\ba 
P(\sigma_T|\bm y_t, \bm w, \bm x_t, \bm u)&=&P(\sigma_T|\bm y_t, \bm w)=\Theta\left(\sigma_T\bm y_t \cdot \bm w \right),\nonumber\\
& & 
\ea 
where $\Theta$ is the step function, the likelihood becomes

\ba 
L(\sigma_e |\bm x,\bm w) & = &\varepsilon(z)\Phi\left(-\bm w \cdot \bm x \sigma_e\right)+(1-\varepsilon(z))\Phi\left(\bm w \cdot \bm x \sigma_e\right)\nonumber\\
& &
\ea 
with the choice $v_p = ||\bm w||^{-2}$  accounting for the effect of inexperience on issue parsing.

These modeling steps allows the partition of the covariance $\bm \Sigma$ to remain block diagonal if it starts block diagonal, where the blocks are the ideological covariance $\bm C_t$ and the affective variance $V_{e|r}$. The evidence 
\ba
Z(\sigma_e|\bm x_t, D_{t})
&=& \E_{Q_{t}}(  L(\sigma_e|\bm u, \bm x_t) )\nonumber \\
&=& \sum_{\sigma_T} \E_z (P(\sigma_e| \sigma_T, \bm x_t, z))\E_{\bm w}(\Theta(\sigma_T\bm w \cdot \bm x_t))\nonumber\\
& &
\ea 
with the expectation taken with respect to the two Gaussian sectors at time $t$. At this point the information concerning the receiver about the past is $D_{t} =\{\hat{\bm w_t}, \bm C_t; \mu_{e|r}, V_{e|r}\}$. 
To reveal interesting symmetries it is useful to introduce the scaled stabilities 
\ba
h_{\bm w}&=&\frac{\hat{\bm w}\cdot\bm x\sigma_e}{\gamma_C}, \,\,\,\,\, h_\mu = \frac{\mu_{e|r}}{\gamma_V}
\label{stabilities}
\ea 
where
\ba
\gamma_C &=& \sqrt{1+\hat {\bm x}\cdot \bm C_t\hat{\bm x}}, \,\,\,\,\,\,
\gamma_V 
= \sqrt{1+V_{e|r}} \label{scales}
\ea 
are the {\it internal response uncertainties}. The evidence depends on the agents state only through the $h_w, h_\mu$ variables: $Z=Z(h_w,h_m)$. The Gaussian integrals lead to

\ba
Z & = &
     \Phi(h_{\bm w}) + \Phi(h_{\mu_{e|r}})  - 2\Phi(h_{\bm w})\Phi(h_{\mu_{e|r}})
\label{eq:evidence}
\ea
where $\Phi(s) = \int_{-\infty}^s\mathrm{d}u g(u)$ is the cumulative of the standard Gaussian $g(u) =(2\pi)^{-1/2}\mathrm{e}^{-\frac{1}{2}u^2}$. 

The learning algorithm, obtained from \ref{eq:evidence},  \ref{dinawvet} and \ref{dinaCvet} is 
\ba
    \hat {\bm w}_{t+1} & =& 
                   \hat {\bm w}_{t} + \frac{1}{\gamma_C}F_w {\bm C_{t}{\bm x}\sigma_e} \label{eq:edw}\\
    \bm C_{t+1} & =&%
              \bm C_{t} + \frac{1}{\gamma_C^2}F_C \bm C_{t}{\bm x}{\bm x}^T\bm C_{t} \label{eq:edc}\\
    \mu_{e|r}(t+1) & = &
                 \mu_{e|r}(t) + \frac{1}{\gamma_V} F_m {V_{e|r}(t)}\label{eq:edm}\\
    V_{e|r}(t+1) & =&
                 V_{e|r}(t) + \frac{1}{\gamma_V^2}F_V {V_{e|r}(t)^2}\label{eq:edv}
\ea
The first two describe the evolution of the ideological sector and the last two of the affective sector. The coupling between the two sets is due to the four $F$ functions  which we now describe.
Note the appearance of the Hebbian-like term $\bm x \sigma_e$ in equation \ref{eq:edw}, which is the effective learning algorithm of the perceptron with a tensorial and adaptive  learning rate modulated by $F_w$. The rest of the dynamics is the adaptive estimation of the noisy channel,
equations \ref{eq:edm} and \ref{eq:edv}. $F_w $ and $F_\mu$  are the {\it Modulation}  functions, since they set the scale of the changes of the parameters of the network for a given example. We also call $F_C$ and $F_V$ modulation functions, which set the scale of changes of the annealing schedule of the learning algorithm.
 These modulation functions $F_w, F_C, F_m$ and $F_V$  are given by
\ba
    F_w (h_w,h_\mu)& =& \frac{\partial\log Z}{\partial h_w} = (1-2\Phi(h_\mu))\frac{g(h_w)}{Z}\label{eq:Fw}\\
    F_C (h_w,h_\mu)& = &\frac{\partial^2\log Z}{\partial h_w^2} = -F_w(F_w + h_w)\label{eq:FC}\\
    F_\mu (h_w,h_\mu)& =& \frac{\partial\log Z}{\partial h_\mu} = (1-2\Phi(h_w))\frac{g(h_\mu)}{Z}\label{eq:Fm}\\
    F_V (h_w,h_\mu)& =& \frac{\partial^2\log Z}{\partial h_\mu^2} = -F_\mu(F_\mu + h_\mu)\label{eq:FV}
\ea
An important feature of equations \ref{eq:Fw} and \ref{eq:Fm} are the prefactors $(1-2\Phi)$ which,  when negative,  allow to learn the opposite of  the arriving information and dynamically lead the society into a group of ferromagnetic or anti-ferromagnetic interactions. A similar reversal of negative
norm-support mechanism was studied in humans in \cite{reversaleffect}.
The fields $h_w, h_\mu$ set the scale of the receiver's opinion on the issue, and its distrust on the emitter, respectively.
\begin{figure}
    \includegraphics[width=1.1\linewidth]{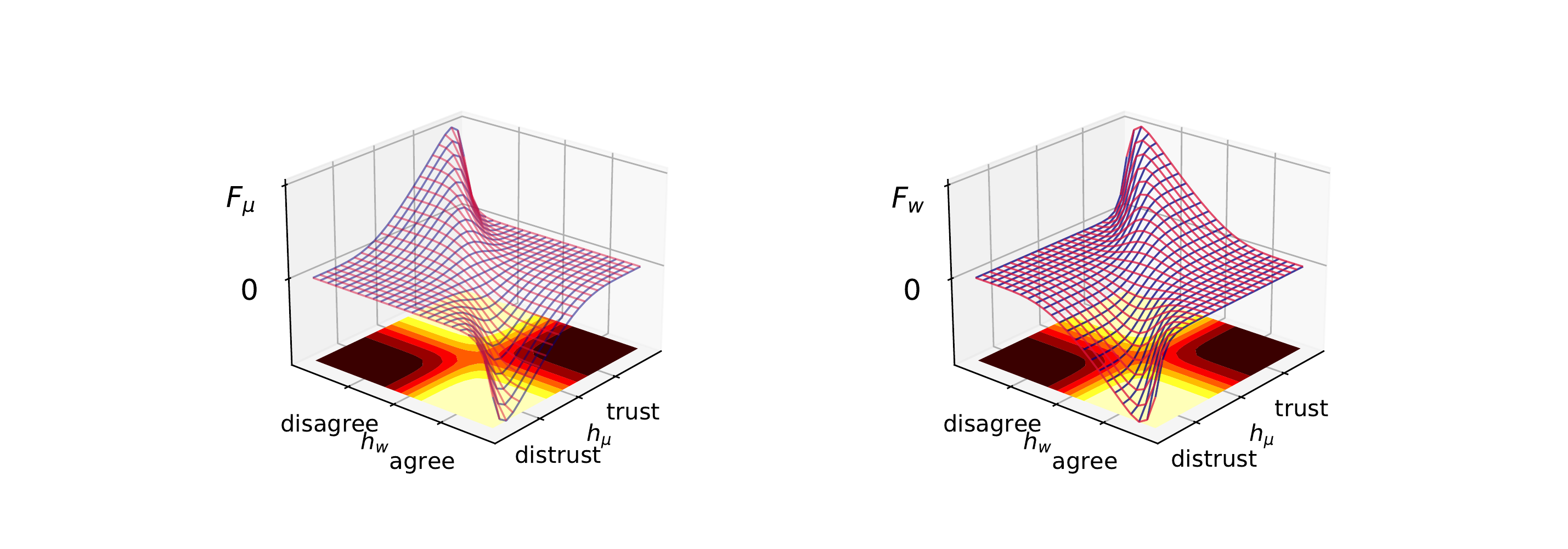}
    \includegraphics[width=1.1\linewidth,left]{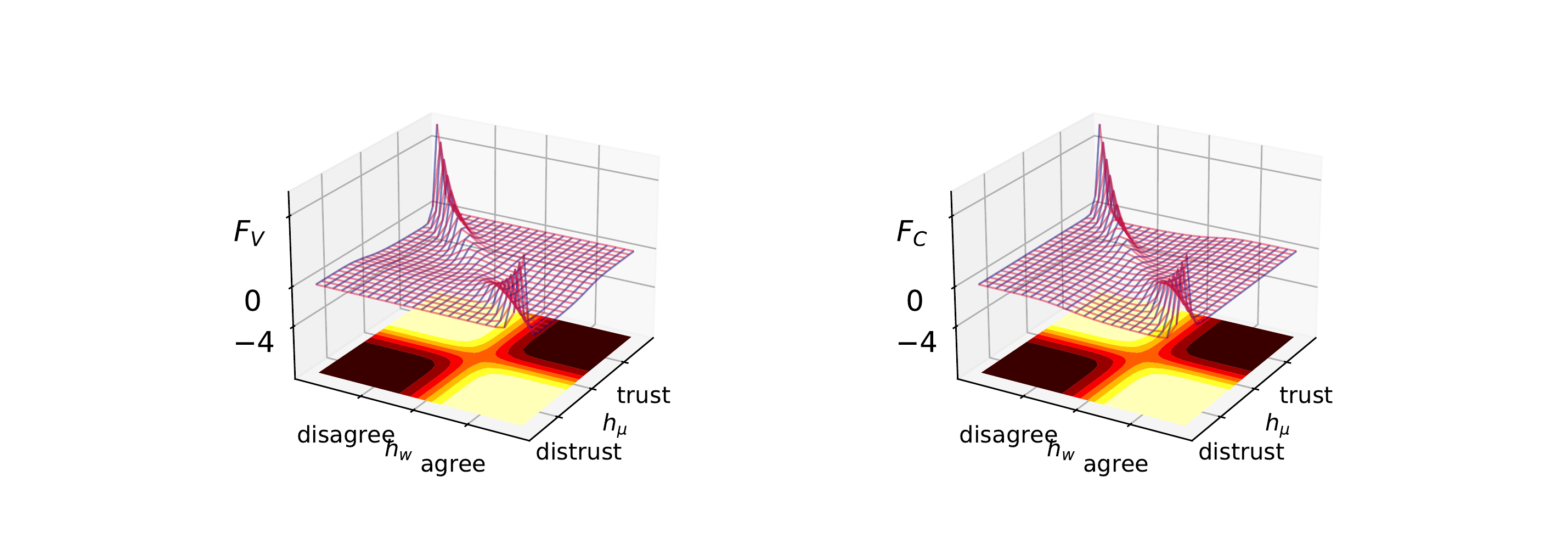}
    \caption{Modulation functions as functions of  disagreement/agreement ($h_w$) and trust/distrust
    ($h_{\mu_{e|r}}$). They are significantly different from zero in the regions of surprises. Top row:  ${ F}_\mu$ for the distrust, left and 
    ${ F}_w$ for  the weights, right. The floor shows a contour graph of the evidence $Z$. The dark regions are where  surprises of agreeing with a distrusted or disagreeing  with a trusted agent occur. Note the complementarity or blame attribution of the two modulation functions. Bottom: Modulation functions for the  covariance modulation functions ${F}_V$ (left, distrust/affinity) and ${ F}_C$ (right, opinion sector). Slightly negative values show the increase in certainty about the distrust and opinion sectors. Large positive values are triggered by surprises and show regions where uncertainty increases. Uncertainties grow along the diagonal $h_w= -h_\mu$. }
    \label{fig:modulation}
\end{figure}
\begin{figure}
    \includegraphics[width=1.1\linewidth,left]{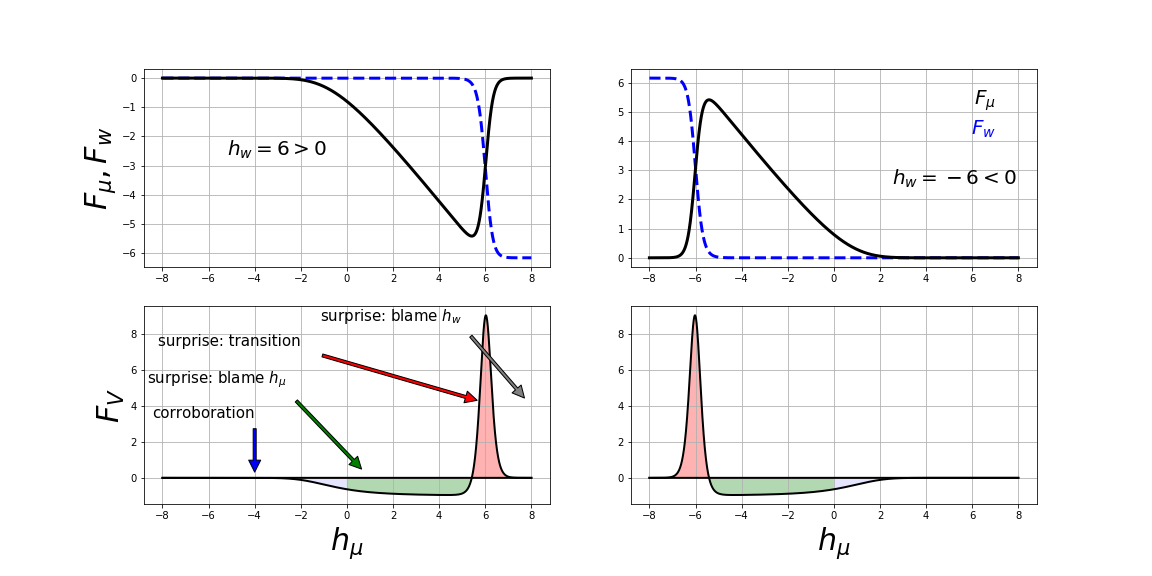}
    \caption{A cut of the functions in figures \ref{fig:modulation} for constant $h_w$. Left: agents agree on an issue, Right: disagree. Top: ${F}_\mu$, continuous black line and ${F}_w$, dashed blue line. Notice the crossover at $h_\mu = h_w$, when the blame for the surprise changes sector. Bottom: ${ F}_V$  shows the increase in uncertainty in the crossover region. }
    \label{fig:modulationcorte}
\end{figure}
From equations \ref{eq:Fw} to \ref{eq:FV}, it is easy to verify  the symmetry between the two sectors, when appropriately scaled
\ba
F_w(h_w,h_\mu) &=& F_\mu(h_\mu,h_w), \\
F_C(h_w,h_\mu) &=& F_V(h_\mu,h_w),
\ea
which have important consequences on the dynamics.

\section{Surprise driven learning\label{algorithmdescription}}
When two agents interact, the receiver undergoes meaningful changes only when there is something that could be interpreted as ``cognitive dissonance" or surprise, see figure \ref{fig:modulation}. The quadrants $h_w h_\mu > 0$ are the regions where the modulation functions are significantly different from zero.
 Agreeing on an issue with a foe occurs for $h_w> 0$ and $h_\mu>0$. When both are negative, the agent disagrees with a trusted agent.  But the modulation functions are not uniform on the surprise region. If $|h_w| > |h_\mu|$, then the affective sector undergoes a large change and the ideological sector remains rather unchanged and vice versa for the reverse inequality. The absolute values  $|h_w| $ and $ |h_\mu|$ are a measure of how sure  the receiver is  of its opinions about the issue ($|h_w|$) or about the emitter ($|h_\mu|$). The sector with the smaller confidence suffers a large change, that tends to eliminate the dissonance.  $h_w$ and $h_\mu$ are fields re-scaled by measures of certainty $\gamma_C$ and $\gamma_V$. Since they are a measure of the width of the prior distribution $Q_t$, they represent a measure of how seriously an agent should take into account its estimates of weights and distrust and are obviously related to Bayesian credible intervals on the sectors. These are also dynamically updated in a complex but intuitive way, shown in 
figure \ref{fig:modulation}. Figure \ref{fig:modulationcorte} shows a cut of $F_\mu$ and $F_w$ (top) and $F_V$ (bottom) for constant $h_w$. Agents disagree $h_w=-6$ on the right, and agree $h_w = 6$ on the left.  
This case illustrates the following features of the modulation functions:
 \begin{itemize}
     \item Region of corroboration, $h_\mu < 0$:     All the modulation functions are effectively zero since there is no surprise (receiver agrees with trusted agent). 
     \item
     For $0\lesssim h_\mu\lesssim 5.5 < h_w$ blame the surprise on the distrust toward the emitter. $F_w$ remains zero since the opinion sector is not blamed for the surprise. Here $F_V$ is negative since the receiver is effectively learning something about the emitter's deserved distrust and thus slightly decreases its uncertainty $V$ ($F_V\lesssim 0$).
     \item Transition surprise: $h_\mu \sim h_w $. There is a surprise and no clear blame attribution since both sectors are similarly confident. $F_\mu$ grows quickly back to zero as $h_\mu$ goes across the transition region and there is a big increase in the uncertainty about the the distrust of the emitter. 
     \item Surprise ($h_\mu > 6.5> h_w$), the receiver blames the opinion sector $h_w$. Once $h_\mu$ is significantly larger than $h_w$, the blame for the surprise falls on the opinion sector and nothing happens on the affective sector. Note that $F_w$ is negative: the receiver learns the opposite of what the emitter is saying.
 \end{itemize}
The agents interaction is driven by the evidence towards minimizing the surprises.
The role of the response uncertainties $\gamma_C $ and $\gamma_V $ (equation \ref{scales}) is to set the proper scale of the effective responses, according to how certain the agent is about the assigned value for that behavior.
Higher values of  $\gamma_C $ or $\gamma_V $ throw the agent's effective response closer to zero, and therefore closer to the surprising zone, so agents who are very sure of their assessments about either their opinion or the distrust they attribute to others are more stubborn. Certainty leads to immunization against surprises.  
The uncertainties also create an asymmetry on the modulation functions regarding novel versus corroborative information, as the less certain an agent is, the smaller is the difference between agreeing or disagreeing with a trusted emitter.
For a given fixed distrust, as an agent becomes more convinced, i.e. smaller norm of the covariance of the Bayesian posterior, surprises become more effective for novel rather than corroborative information. A possible link between the width of the posterior, its role in weighting the importance of surprises and possible political orientations of agents in a small political agenda context is discussed in \cite{CaVi2011}, \cite{CaCeVi2015}.

Equipped with the full description of EDNNA for perceptron agents, we proceed with the study of societies of EDNNA agents learning through communication.

\section{A society of $N$ agents\label{Nagents}}

Consider a society of $N$ agents,  exchanging $\pm 1$ opinions about $P$  issues $ \bm x \in \R^K$, independently chosen with $|\bm x|=1$. Different levels of frustration, polarization and dynamical annealing can result depending on the parameters and the initial conditions. In order to control  the effect of the initial conditions we need to parametrize them in an appropriate manner.

We start with agents, with weight vectors $\bm w_i \in \R^K$, chosen uniformly at random independently from anything else, on the sphere of radius $|\bm w| = \sqrt{1+c}$. For each agent the initial  co-variance matrix is the same $\bm C= c\bm I$. Each agent $i$ has an initial set of distrusts $\mu_{j|i}$ towards agents $j\neq i$ drawn from a normal distribution $\mu_{j|i} \sim G(\mu_0,\sqrt{1+V_0})$. These choices permit comparing the typical changes  $\langle|\Delta w|\rangle$ e $\langle|\Delta \mu|\rangle$, as initially $\langle h_{\bm w}\rangle \approx \langle h_{\mu}\rangle$.
If all agents trust every other agent, $\mu_0 \ll 0$, the system typically evolves toward consensus, with the overlaps $\rho_{ij} = \frac{\bm w_i \cdot \bm w_j}{|\bm w_i ||\bm w_j|} \lesssim 1$, so we will not deal with this trivial case.

\begin{figure}[ht]
\includegraphics[width=.5\linewidth]{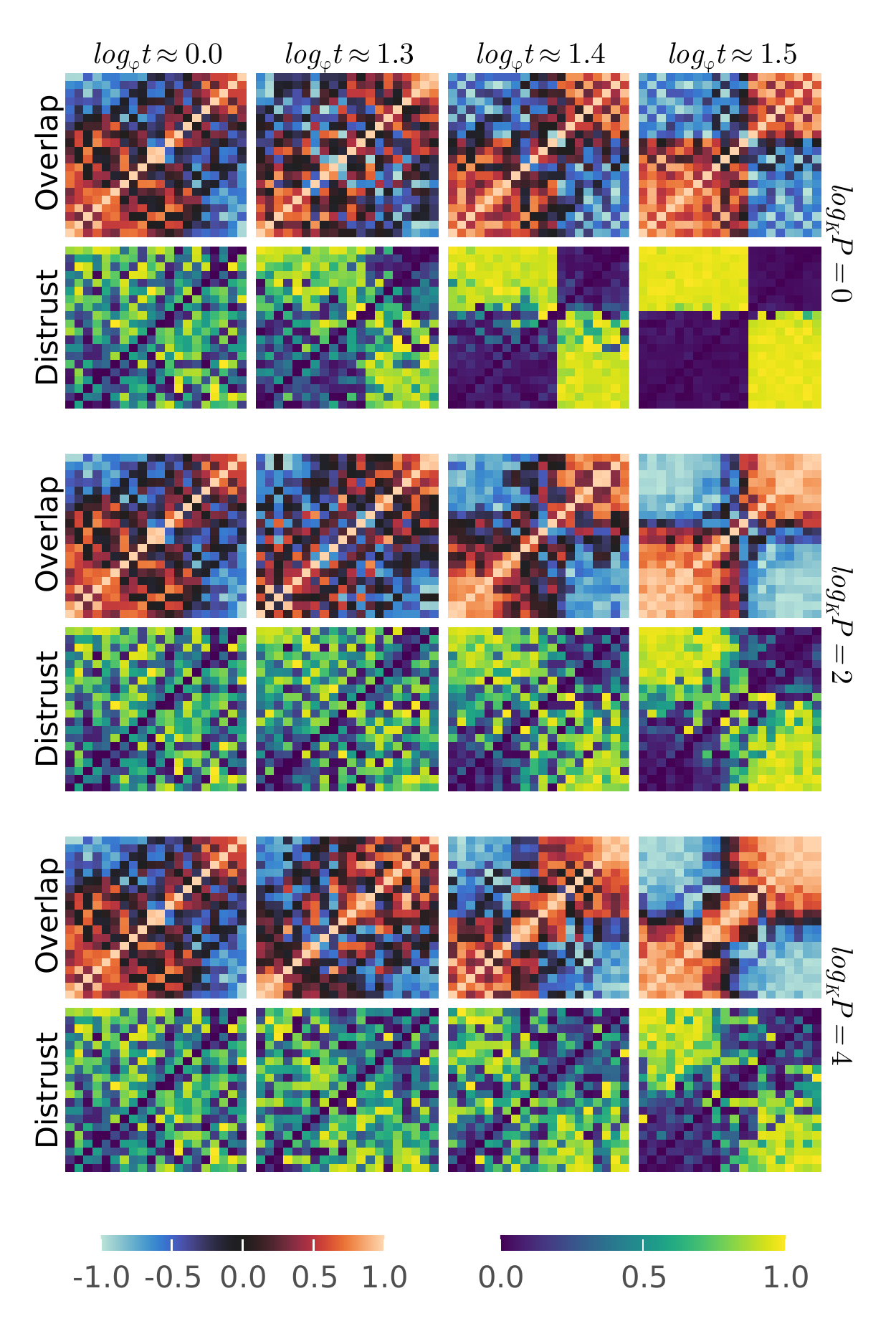}
\caption{ Evolution of the overlap and distrust matrices for a society with $N=20$ agents with $K=5$ ideological space dimensions, full covariance matrix with initial norms $C_0 = 1.0$ and $V_0 = 1.0$ for different number of issues. As the number of issues increase, the polarization goes from affective to ideological. $\varphi = N(N+K-1) =480$ is the number of (effectual) degrees of freedom.}
\label{matrices-consensus}
\end{figure}
\begin{figure}[ht]
\includegraphics[width=.7\linewidth]{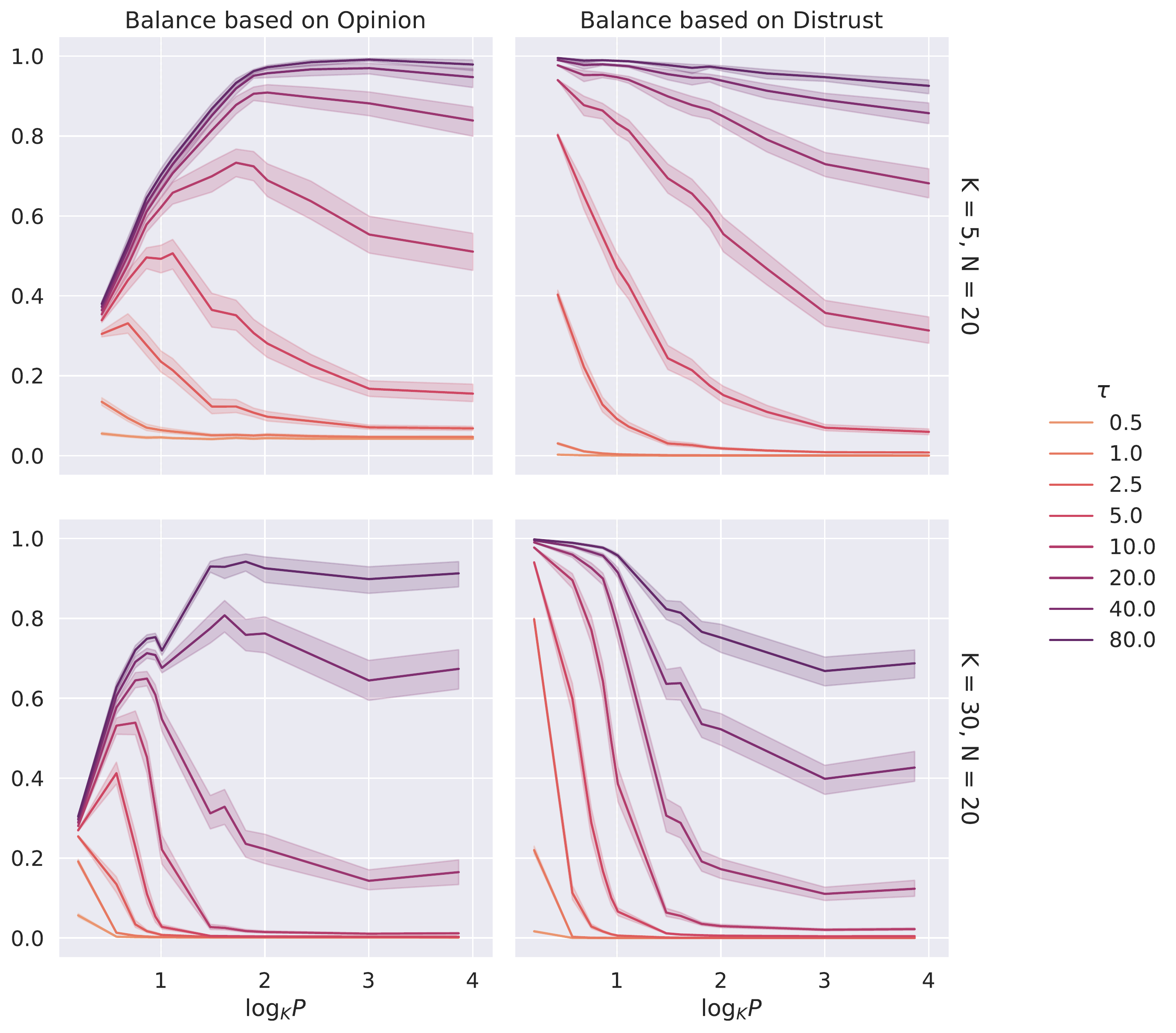}
\caption{ The mean value of the balance distribution describes the frustration based on ideology/opinion $B_I$ (left) and on affinity/distrust $B_A$ (right); for small (upper) or large (lower) dimension of issues compared to the size of the population, as a function of the number of issues under discussion, for different simulation times $\tau = \frac{t}{\varphi} =\frac{t}{N(N + K - 1)}$. The agents evolve with a full covariance matrix $\bm C$. }
\label{balance01}
\end{figure}

\begin{figure}[ht]
\includegraphics[width=\linewidth]{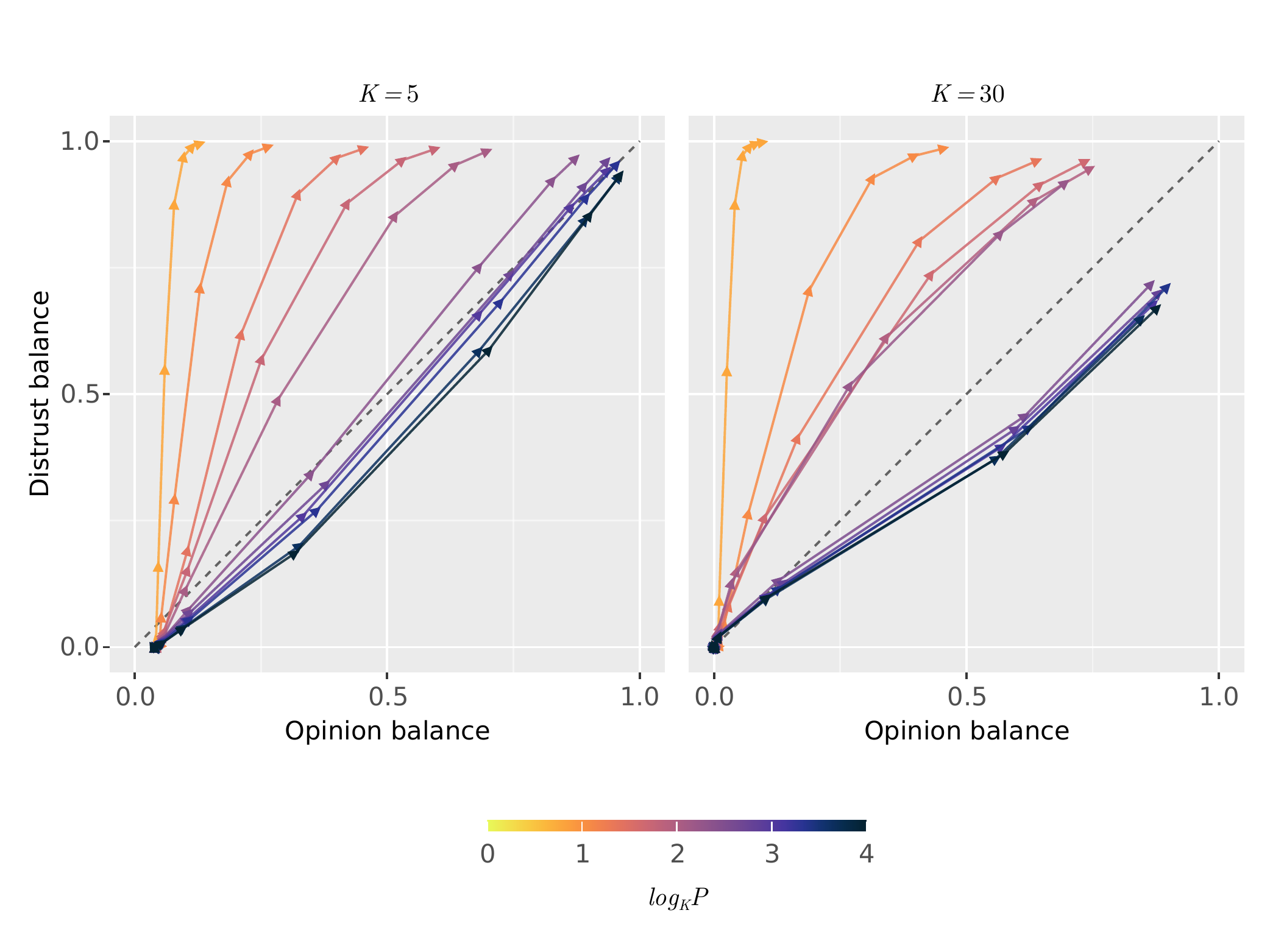}
\caption{ Time flow for $B_A$ versus $B_I$ for different sizes of the agenda. Same data as in figure \ref{balance01}. Full covariance case. For simple agendas affinity polarization sets in before ideological polarization. This is reversed when  larger sets of issues are under discussion.}
\label{balance-flow-matrixcov}
\end{figure}

\begin{figure}[ht]
\includegraphics[width=.7\linewidth]{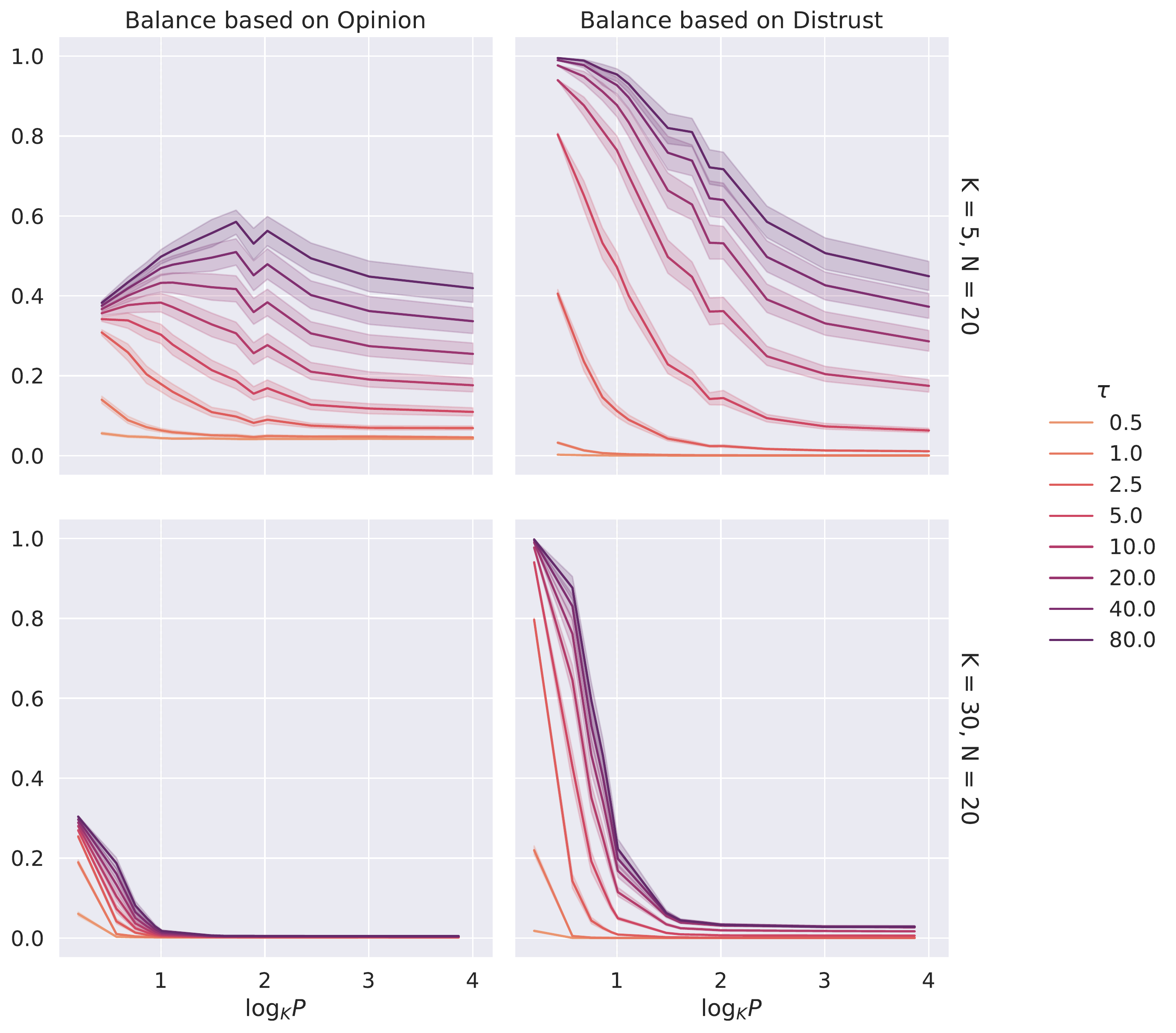}
\caption{ Same as figure \ref{balance01}, but the agents evolve with a covariance matrix that is proportional to the identity $\bm C= c\bm I$. }
\label{balance02}
\end{figure}

\begin{figure}[ht]
\includegraphics[width=\linewidth]{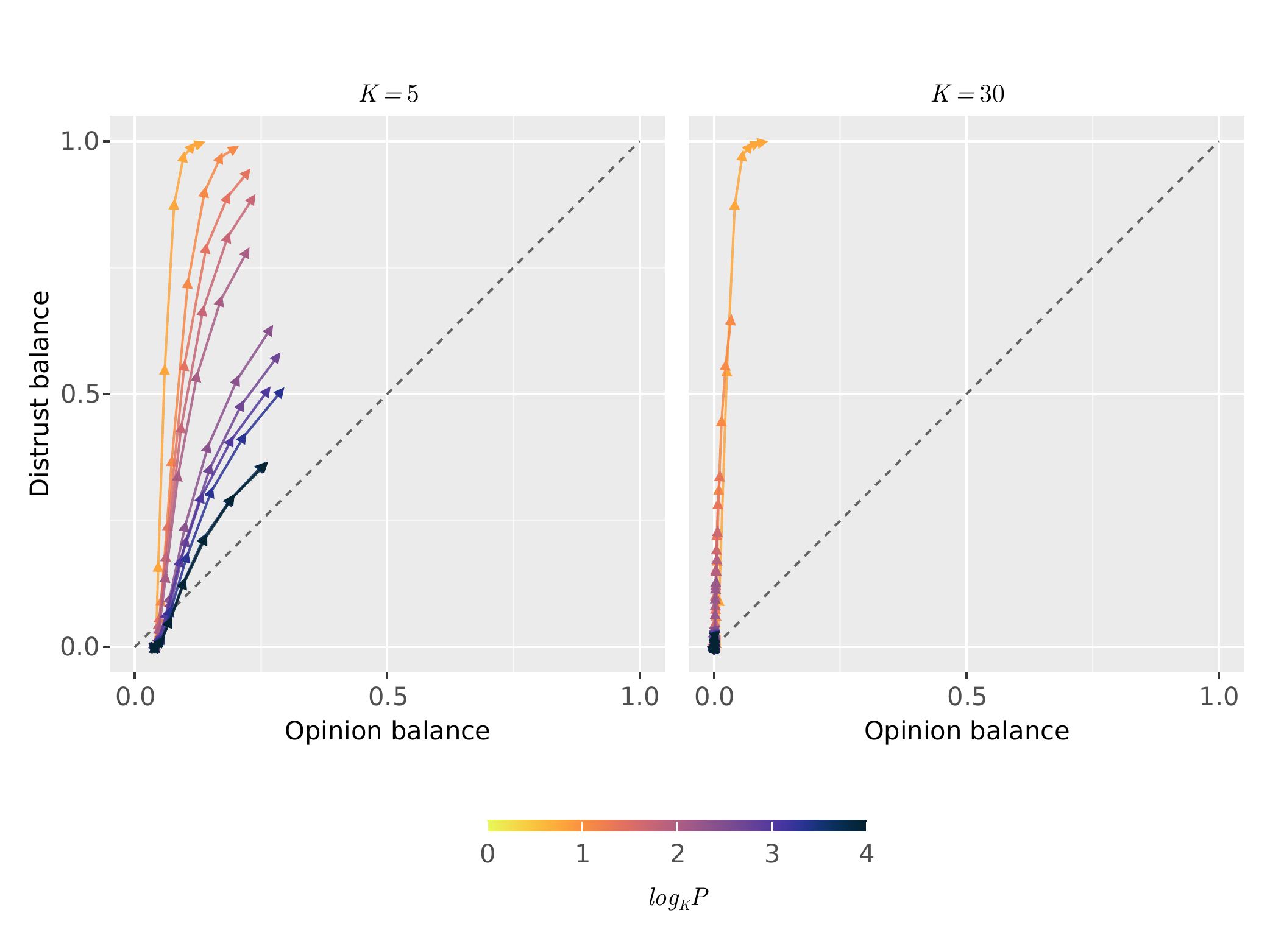}
\caption{ Time flow of $B_A$ versus $B_I$ for different sizes of the agenda. Same data as in figure \ref{balance02}. Simple covariance case $\bm C= c\bm I$. For these simpler agents glassy behavior persists longer and  affinity polarization sets in before ideological polarization. }
\label{balance-flow-scalarcov}
\end{figure}

To characterize the state of the ideological and affective sectors, \cite{KlarAffective} \cite{IyengarAffective}, we introduce order parameters that describe the balance of
trust \cite{Heider}  \cite{HarariBalance} and balance of opinions in triads of agents.
Receiver $i$, which distrusts emitter $j$ by a scaled distrust $h_\mu$, communicates with an information channel with flipping probability $\varepsilon_{j|i} = \Phi(h_\mu)$. Define $\upsilon_{j|i} = 1-2\varepsilon_{j|i}$, which is positive/negative for trust/distrust. 
For any three agents, following \cite{Heider}, we say  that the trust relation is balanced 
if  $b_{ijk} := \upsilon_{j|i}\upsilon_{k|j}\upsilon_{i|k} > 0$. This is as frustration in spin glasses but for directed graphs since the odd permutation $b_{ikj}$ may be of opposite sign. The affective balance  is the population average
\ba
B_A &=& \frac{1}{2N_T}\sum_{\langle ijk\rangle}(b_{ijk}+b_{ikj}),
\ea 
characterizes the state of affective frustration of a society. The sum runs over the set of $N_T= N$ choose $3$ triplets.
In addition to frustration associated to breakdown of affective balance, a second type of frustration can be defined.
The opinion alignment is measured by the symmetric overlap $\rho_{ij}$. We can measure opinion balance by introducing for every triplet of agents the ideological balance and characterize the state by the average over the population
\ba
B_I &=& \frac{1}{N_T}\sum_{\langle ijk\rangle}\rho_{ij}\rho_{jk}\rho_{ki}.
\ea 
A ferromagnetic state or a staggered polarized society will have a $B_I$ and $B_A$ close to one; values for a highly frustrated society will be close to zero. These quantities may anneal to $1$, but if the time scale is very large it points to a society that perseveres in a spin glass like state, without the need to have an infinite separation of time scales. 

For agents using the tensorial learning adaptive EDNNA algorithm, with a full covariance $\bm C$, a pattern emerges as the number $P$ of issues is varied, see figures \ref{matrices-consensus}, \ref{balance01} and \ref{balance-flow-matrixcov}. For a few issues the society polarizes rapidly into two factions in the affective sector and then more slowly the affective polarization drives the ideological polarization. However as the agenda grows in complexity the ideological polarization sets in before and then drives the affinity polarization. This is seen in the dark lines below the diagonal in figure \ref{balance-flow-matrixcov}. Times to anneal into the polarized state grow and the society lingers in a spin-glass state, with small $B_A$, thus large frustration. 
 The time to achieve a balanced society increases as $P$ increases. Times $\tau$ measures the  number of learning interactions per degrees of freedom of the system ($\varphi = N(N+K-1)$) so that we can compare simulations with different parameters. 
This reversal disappears when the agents use a simpler algorithm, with the covariance $\bm C = c\bm I$, see figures  \ref{balance02}, \ref{balance-flow-scalarcov}. For these simpler agents, affinity polarization always seem  to occur first and  has slow  behavior annealing form the glassy state to the  ideological polarized state. 

\section{Discussion and Conclusions}
Evolution under certain conditions leads to learning algorithms adapted to that environment. Our agents use algorithms designed to be efficient in a small group scenario, more specifically where there is a rule to be  learned  and followed by members of this group. This gives rise to the opinion sector, with a mechanism to emit opinions about issues.  The interaction with other agents and the possibility of noisy communications or concealed cheating demands that the optimized algorithm incorporates a defense mechanism that appears in the assignment of a level of distrust to other members of the group. The entropic dynamics for NN architectures (EDNNA) analysis provides a general method to obtain such optimized  learning algorithm. It could also  result of an evolutionary process, similar to that shown in \cite{Neirotti2003}. Once a two agent interaction is defined by the exchange and learning of issues with these rich algorithms, we construct a society.   The reader should not be induced to think that we have obtained {\it the} Bayesian algorithm. There is no such thing. There are Bayesian algorithms conditioned on some informational context. Failure to understand this has lead to claims that humans' lack of rationality can't be modeled by Bayesian methods \cite{EberhardtDanksproblematic}, but the problem in modeling humans is not that information theory does not apply but that the informational structure is poorly determined in the model.  
A simpler covariance (multiple of the identity) model,  such as in figure \ref{balance-flow-scalarcov} leads to algorithms just as Bayesian, but under a different set of constraints.

The society of $N$ agents has, as expected, a great variety of possible behaviors. The interesting result we report is that if the agenda is simple opposing parties can form around polarized affinities which proceed and push on a slower time scale to shared ideologies. However as it grows in complexity (larger $P$) the time to anneal down frustration grows and unbalanced societies have an effective spin-glass regime in finite times that might be much larger that the life of an agent, and thus seems to evolve in the presence of quenched disorder.
Now, the ideological  sector anneals first,  driving the slower  affinity polarization.  The reduction of  times to polarize that accompanies the  reduction of the agenda is in accordance with examples discussed in \cite{FiorinaPolarization} where rapid partisan polarization is not accompanied by rapid changes in positions on economic policy issues. But this doesn't happen for the simpler scale covariance agents. Affinity polarization sets in first and ideological polarization comes later at a much slower scale.

Agents with information processing capabilities  permit a general framework  useful in  different situations. 
Questionnaires are a central method to gather information in the humanities and these NN agents can include   such scenarios \cite{CaVi2011}. The different sectors are natural from an information processing formulation, lead to different frustrations and polarizations which are compatible with those observed in political science and permit sorting conditions when affective polarization leads ideological polarization \cite{IyengarAffective}. We have used mistrust as a portmanteau for several associated concepts \cite{YamagishiTrust} \cite{InterdisciplinaryPerspectivesonTrustSchockleyEdit}, since their nuances are  beyond our capacity. The NN we use are certainly simple and limited, despite the rich learning algorithms they use. The EDNNA formalism applies to any architecture, however it becomes intractable fast with architecture complexity.  This of course is just an obvious reminder that  care should be taken in building metaphors that map simple ideas in societies of NN to human societies.

{\bf Acknowledgment:} We thank A. Caticha, O. Kinouchi, R. Vicente, M. Copelli, JP Neirotti  for discussions. FA received financial support from a  
Conselho Nacional de Desenvolvimento Científico e Tecnológico
 (CNPq) PhD fellowship. This work received partial support from CNAIPS-NAP USP.

\bibliography{EDNNA_frust.bib}

\end{document}